\documentclass[11pt]{article}
\usepackage{latexsym,amssymb,graphicx,natbib,amsmath,eufrak,color,verbatim,soul}
\usepackage[latin1]{inputenc}

\title{Dealing with non-ignorable nonresponse in survey sampling:\\ a latent modeling approach}

\author{Alina Matei\thanks{Institute of Statistics, University of Neuch\^atel,
Pierre à Mazel 7, 2000, Neuchâtel, Switzerland,
alina.matei@unine.ch and Institute of Pedagogical Research and
Documentation Neuchâtel, Switzerland} \and M. Giovanna Ranalli\thanks{Dept. of Economics, Finance and Statistics, University of Perugia, Italy, giovanna.ranalli@stat.unipg.it}}

\def\2xk{{\bf x}_{k2}}
\def\1xk{{\bf x}_{k1}}
\def\0xk{{\bf x}_{k0}}

\def\b#1{\mbox{\boldmath $#1$}}    
\def\m#1{\mbox{#1}}                
\def\ml#1{\mbox{\scriptsize #1}} 
\def\ha#1{\mbox{$\b{\widehat{ #1}}$}}

\begin{document}

\baselineskip 22 pt
\maketitle

\begin{abstract}
Nonresponse is present in almost all surveys and can severely bias
estimates. It is usually distinguished between unit and item
nonresponse. By noting that for a particular survey variable, we
just have observed and unobserved values, in this work we exploit
the connection between unit and item nonresponse. In particular,
we assume that the factors that drive unit response are the same
as those that drive item response on selected variables of
interest. Response probabilities are then estimated using a latent
covariate that measures the \emph{will to respond to the survey}
and that can explain a part of the unknown behavior of a unit to
participate in the survey. This latent covariate is estimated
using latent trait models. This approach is particularly relevant
for sensitive items and, therefore, can handle non-ignorable
nonresponse. Auxiliary information known for both respondents and
nonrespondents can be included either in the latent variable model
or in the response probability estimation process. The approach
can also be used when auxiliary information is not available, and
we focus here on this case. We propose an estimator using a
reweighting system based on the previous latent covariate when no
other observed auxiliary information is available. Results on its performance are encouraging from simulation studies on both real and simulated data. 

~
\end{abstract}

\section{Introduction}

Nonresponse is an increasingly common problem in surveys. It is a
problem because it causes missing data and, more importantly,
because such gaps are a potential  source of bias for survey
estimates. In the presence of unit nonresponse, it is often
assumed that each unit in the population has an associated
probability to respond to the survey. Such a response probability
is unknown and several methods are proposed to estimate it either
explicitly, using response propensity modeling like logistic
regression models \citep[see e.g.][]{kim:kim:07}, or implicitly,
using response homogeneity groups or more generally calibration
\citep[see][for an overview]{sar:lun:05}. Once estimates are
computed, a commonly used method to deal with unit nonresponse is
reweighting: sampling weights of the respondents are adjusted by
the inverse of the estimated response probability providing new
weights. Estimation of response probabilities typically requires
the availability of auxiliary information, either in the form of
the value of some auxiliary variables for all units in the
originally selected sample or of their population mean or total.

In this paper, we are particularly interested in the case where
the missing data mechanism is non-ignorable, because nonresponse
depends on characteristics of interest that are either observed
only on the respondents or are completely unobserved, which leads
to data that are Not Missing At Random (NMAR). This is typical of,
but not limited to, surveys with sensitive questions (concerning
drug abuse, sexual attitudes, politics, income etc). Various
approaches are proposed in the survey sampling literature to deal
with non-ignorable nonresponse. These approaches can be roughly
divided into likelihood based methods and reweighting methods.
Note that all of these methods make use of observed auxiliary
information. Survey problems with non-ignorable nonrespondents are
discussed e.g. in \citet{gre:ree:zie:82}, \citet{lit:rub:87},
\citet{beu:00}, \citet{qin:leu:sha:02}, \citet{zha:02}.
\citet{cop:far:98} introduce into the British National Survey of
Sexual Attitudes and Lifestyles a variable called
`enthusiasm-to-respond' to the survey, which is expected to be
related to probabilities of unit and item response. A method is
proposed that estimates these probabilities using this variable to
achieve unbiased estimates of population parameters. An approach
based on the use of latent variables for modeling nonignorable
nonresponse is given in \citet{bie:lin:07}, extending the ideas in
\citet{dre:ful:80} and using a discrete latent variable based on
call history data available for all sample units. The latent
variable is computed using some indicators of level of effort
based on call attempts.

We propose here a method of reweighting to reduce nonresponse bias
in the case of non-ignorable nonresponse. The method does not
require the availability of auxiliary information, on the sample
or population level, but different assumptions are made. First, it
is assumed that item nonresponse is present in the survey and that
it affects $m$ variables of particular interest. Thus a response
indicator can be defined for each variable $\ell$, for
$\ell=1,\ldots,m$, taking value 1 if item $\ell$ is observed on
unit $k$ and 0 otherwise. Next, the response indicators are
assumed to be manifestations of an underlying continuous scale
which determines a latent variable that is related to the response
propensity of the units and to the variable of interest. It is
possible to compute such a latent variable for all units in the
sample, not only for the respondents, and thus to use it as an
auxiliary variable in a response probability estimation procedure.
The outcome of this estimation procedure can finally be used in a
reweighting fashion.

The use of  continuous latent variables to model item nonresponse
is considered in \citet{mou:kno:00}. In this paper, we take a
different perspective and use latent variable models to address
non-ignorable unit nonresponse. We propose to use a latent
variable called here `will to respond to the survey', which is
expected to be related to the probability of unit response,
similar to the case of the `enthusiasm-to-respond' variable as
defined by \citet{cop:far:98}. Following \citet{mou:kno:00},
`\emph{weighting through latent variable modeling is expected to
perform well under non-ignorable nonresponse where conditioning on
observed covariates only is not enough}.' Moreover, in the absence
of any covariate, we expect that an estimator based on the
proposed weighting system using latent variables will perform
better in terms of bias reduction than the naive estimator
computed on the set of respondents. \citet{mou:kno:00} propose a
reweighting system for \emph{item non-response} using covariates
and one or more latent variables. Our major contribution over the
existing literature is to construct a weighting system to deal
with \emph{unit and item non-response} based only on latent
variables and that can also be used in the absence of any other
covariate. On the other hand, our approach is  different to that
of \citet{cop:far:98}, because they survey their
`enthusiasm-to-respond' variable on the respondents to quantify
the interest in answering the survey and a set of covariates,
while we infer it from the data.

The paper is organized as follows. Section \ref{sec:frame}
introduces the survey framework and notation. Section 3
illustrates estimation of response probabilities. Section
\ref{sec:latmod} describes the latent trait model used to this
end. The proposed estimator and its variance estimation are shown
in Section \ref{estimation}. In Section \ref{sec:MCsim}, the
empirical properties of the proposed estimator are evaluated via
simulation studies. In Section \ref{sec:concl} we summarize our
conclusions.

\section{Framework}\label{sec:frame}

Let $U$ be a finite population of size $N$, indexed by $k$ from
$1$ to $N.$ Let $s$ denote the set of sample labels, so that
$s\subset U,$ drawn from the population using a probabilistic sampling design
$p(s)$. The sample size is denoted by $n.$ Let $\pi_k=\sum_{s;
s\ni k} p(s)$ be the probability of including unit $k$ in the
sample. It is assumed that $\pi_k>0$, $k=1, \dots, N.$ Not all
units selected in $s$ respond to the survey. Denote by $r\subseteq
s$ the set of respondents, and by $\bar{r}=s\setminus r$ the set
of nonrespondents. The response mechanism is given by the
distribution $q(r|s)$ such that for every fixed $s$ we have
$$q(r|s)\geq 0, \mbox{ for all } r\in \mathcal{R}_s \mbox{ and }
\sum_{s\in {\mathcal R}_s} q(r|s)=1, \mbox { where } {\mathcal R}_s=\{r |
r \subseteq s\}.$$

Under unit nonresponse we define the response indicator $R_k=1$ if
unit $k\in r$ and 0 if $k \in \bar{r}$. Thus $r=\{k \in s | R_k=1\}.$ 
We assume that these random variables are independent of one
another and of the sample selection mechanism \citep{oh:sch:83}.
Since only the units in $r$ are observed, a response model is used
to estimate the probability of responding to the survey of a unit
$k\in U,$ $p_k=P(k\in r| k\in s)=P(R_k=1| k\in s)$, which is a
function of the sample and must be positive.

Suppose that in the survey there are $m$ variables of particular
interest. Each respondent is exposed to these $m$ questionnaire
variables, labelled $\ell=1, \dots, m.$ Suppose that the goal is
to estimate the population total of some variables of interest
and, in particular, of the variable of interest $y_j$, i.e.
$Y_j=\sum_{k=1}^N y_{kj}$, with $y_{kj}$ being the value taken by
$y_j$ on unit $k$. In the ideal case, if the response distribution
$q(r|s)$ is known, then the $p_k$'s would be known and available
to estimate $Y_j$ using a reweigthing approach. Suppose also that
item nonresponse is present for variable $y_j.$ Let $r_j=\{k
\mbox{ answers } y_j | k\in r\}$ be the set of respondents for
variable $y_{j}.$ As in the case of unit nonresponse we assume
that the units in $r_j$ respond independently of each other. Let
$q_{kj}=P(k \mbox{ answers } y_j  | k\in r).$ The final set of
weights to be used into a fully reweighting approach to handle
unit and item nonresponse is given by $1/(\pi_k p_k q_{kj}),$  for
all $k\in r_j,$ assuming $q_{kj}>0.$ These weights can be for
example used in a three-phase fashion in the following
Horvitz-Thompson (HT) estimator
\begin{equation}\label{est1}
\widehat{Y}_{j,pq,\ml{true}} = \sum_{k\in r_j}\frac{y_{kj}}{\pi_k
p_kq_{kj}},
\end{equation}
\citep[see][for the properties of estimators under three-phase sampling]{legg2009two}.

Usually, $p_k$ and $q_{kj}$ are unknown and should be estimated. A
nonresponse adjusted estimator is then constructed by replacing $p_k$
and $q_{kj}$ with estimates $\widehat{p}_k$ and $\widehat{q}_{kj}$
in (\ref{est1}). The following sections provide details with this regard.

\section{Estimating response probabilities} \label{sec:estp}

\subsection{Using logistic regression to estimate $p_k$}

Different methods to estimate ${p}_k$ are proposed in the
literature. All of these methods are based on the use of auxiliary
information known on the population or sample level. In the case
of non-ignorable nonresponse,  the variable of interest is itself
the cause (or one of the causes) of the response behavior, and a
covariance between the former and the response probability is
produced through a direct causal relation \citep[see][]{gro:06}.
In such a case, the response probability $p_k$ could be modeled
for $k\in s$ using logistic regression as follows
\begin{equation}\label{nonig1}
p_k=P(R_k=1| y_{kj})=\frac{1}{1+\exp(-(a_0+a_1y_{kj}))},
\end{equation}
or as follows
\begin{equation}\label{nonig2}
p_k=P(R_k=1| y_{kj},
\mathbf{z}_k)=\frac{1}{1+\exp(-(a_0+a_1y_{kj}+{\mathbf{z}_k'\boldsymbol\alpha}))},
\end{equation}
where $\mathbf{z}_k=(z_{k1}, \dots, z_{kt})'$ is a vector with the
values taken by $t\geq 1$ covariates on unit $k$, and $a_0$, $a_1$
and $\b{\alpha}$ are parameters.

Nonresponse bias in the unadjusted respondent total of the
variable of interest $y_j$ depends on the covariance between the
values $y_{kj}$ and $p_k$ \citep[see][]{bet:88}. An example of a
covariate that reduces the covariance between $y_{kj}$ and $p_k$
is the interest in the survey topic, such as knowledge, attitudes,
and behaviors related to the survey topic
\citep[see][]{gro:all:06}. The set of covariates $\mathbf{z}_k$
could be also related to the variable of interest $y_{j}$ to
reduce sampling variance \citep{lit:var:05}.

Since $y_{kj}$ is only observed on respondents, Models
(\ref{nonig1}) and (\ref{nonig2}) cannot usually  be estimated. Therefore,
usually, the values of $\mathbf{z}_k$ that are known for both
respondents and nonrespondents  and are related to the $y_{kj}$'s
by a `hopefully strong regression' \citep{cas:sar:wre:83} are used
in the following   model
\begin{equation}\label{nonig33}
p_k=P(R_k=1| \mathbf{z}_k)=\frac{1}{1+\exp(-(a_0+\mathbf{z}_k'{\boldsymbol{\alpha}}))}.
\end{equation}
Then,  maximum likelihood  can be used to fit Model
(\ref{nonig33}) using the data $(R_k, \mathbf{z}_k)$ for $k\in s$.
This leads to estimate $\widehat a_0$ and
$\widehat{\boldsymbol{\alpha}}$  and to the estimated response
probabilities $ \widehat{p}_k=1/(1+\exp(-(\widehat
a_0+\mathbf{z}_k'\ha{\alpha}))$ to be used in (\ref{est1}). This
procedure provides some protection against nonresponse bias if
$\mathbf{z}_k$ is a powerful predictor of the response probability
and/or of the variable of interest \citep{kim:kim:07}.

In what follows, we propose a reweighting adjustment  system based
on an auxiliary variable that measures the propensity of each unit
to participate to the survey. To this end, further assumptions on
the response model are introduced in order to assume a dependence
of the $p_k$'s on one latent auxiliary variable that is connected
to the propensity scores of \citet{ros:rub:83}. The proposed
approach can be used when no other auxiliary information is
available on $k\in s$.

\subsection{Latent variables as auxiliary information}

To obtain a measure of  response propensities, we consider the
case in which item nonresponse on the variables of interest is
also present. Then, following \citet[p.278]{cha:ski:03}:
`\emph{from a theoretical perspective the difference between unit
and item nonresponse is unnecessary. Unit nonresponse is just an
extreme form of item nonresponse}', we assume that item response
on the variables of interest is driven on respondents by the same
attitude and factors that drive unit response. Latent variable
models can be used to estimate such factors that, therefore, can be used as
covariates in a logistic response model.

As we have already mentioned we assume that item nonresponse
affects the $m$ survey variables of particular
interest. A second response indicator is introduced for each item
$\ell$. For each item $\ell$ and each unit $k$, a binary variable
$x_{k\ell}$ is defined that takes value 1 if unit $k$ answers to
item $\ell$ and $0$ otherwise. Let $\mathbf{x}_k=(x_{k1}, \dots,
x_{k\ell}, \dots, x_{km})'$ denote the vector of response
indicators for unit $k$ to the $m$ items and let
$\mathbf{y}_k=(y_{k1}, \dots y_{k\ell}, \dots, y_{km})'$ be the
study variable vector for unit $k.$ Thus $y_{k\ell}$ is the
response value of unit $k$ to item $\ell$ and $x_{k\ell}$ is its
response indicator.

Suppose the $x_{k\ell}$'s are related to an assumed underlying latent
continuous scale; they are the indicators of a latent variable
denoted by $\theta_k.$ \citet{dem:bar:96} call the variable
$\theta_k$ the `tendency to respond' to the survey.  We call it here
the `will to respond to the survey' of unit $k.$ A latent trait model with a single latent variable is used to
compute $\theta_k$ for each $k\in s$ (we will see later how; see
Section \ref{sec:esttheta}). Assume for the moment that $\theta_k$ is known on all sample units
and, as with usual auxiliary information,  can be used as a
covariate. In the absence of other covariates, Model (\ref{nonig33}) is
rewritten as
\begin{equation}\label{nonig22}
p_k=P(R_k=1|\theta_k)=\frac{1}{1+\exp(-(\alpha_0+\alpha_1\theta_k))}.
\end{equation}
The covariate $\theta_k$ can be viewed as a variable explaining the
behavior related to the survey topic, and thus having good
properties to reduce the covariance between $y_{kj}$ and $p_k$ and
the nonresponse bias. If other suitable auxiliary
information is available, it can be inserted in the model as
supplementary covariates. Now, to estimate the parameters of Model
(\ref{nonig22}), the value of $\theta_k$ has to be available for
all units in the sample. The following sections provide details on
how to obtain estimated values of $\theta_k$ for both respondents
and nonrespondents.

\section{Computing response propensities using latent trait models}\label{sec:latmod}
The variable $\theta_k$ can be computed using a latent trait model.
In general, latent variable models are multivariate regression
models that link continuous or categorical responses to unobserved
covariates. A latent trait model is a factor analysis model for
binary data \citep[see][]{bar:all:02, skr:rab:07}.

\begin{figure}[htb]
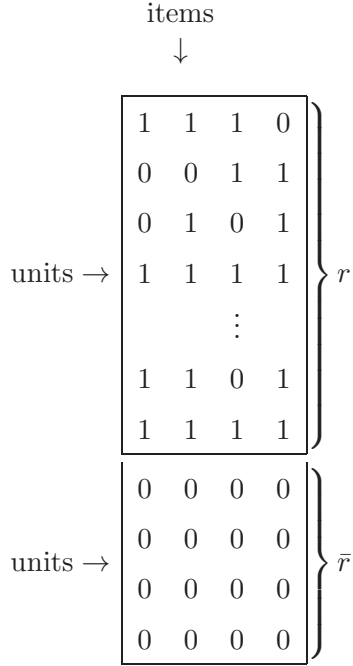

\begin{center}
items  \\
$\downarrow$
$$
\mbox{units} \rightarrow \left.
\begin{tabular}{|cccc|}
\hline
1 & 1 & 1 & 0 \\
0 & 0& 1 & 1 \\
0 & 1 & 0 & 1 \\
1 & 1 & 1 & 1 \\
& & $\vdots$ & \\
1 & 1 & 0 & 1 \\
1 & 1 & 1 & 1 \\
\hline
\end{tabular}
\right\} r
$$
\vspace*{-0.35 cm}
$$
\mbox{units} \rightarrow \left.
\begin{tabular}{|cccc|}
0 & 0 & 0 & 0 \\
0 & 0 & 0 & 0 \\
0 & 0 & 0 & 0 \\
0 & 0 & 0 & 0 \\
\hline
\end{tabular}
\right\} \bar{r}
$$
\end{center}
\caption{Schematic  representing variables $x_{k\ell}$ for the sets $r$
and $\bar{r}$ } \label{figure1}
\end{figure}

We start by creating the matrix with elements $\{x_{k\ell}\}_{k\in
s; \ell=1,\ldots,m}$. Figure \ref{figure1} shows a schematic of
the indicators $x_{k\ell}$ for respondents and nonrespondents.
Then, we assume that the factors that drive unit response are the
same as those that drive item response on selected variables of
interest. In other words, item nonresponse is assumed
nonignorable.

Let
$q_{k\ell}$ be the probability of response of unit $k$ for item
$\ell,$ for all $\ell=1, \dots, m$ and $k\in r.$ As in the case of
unit nonresponse, $q_{k\ell}$ is modelled as a function of the
variable of interest using logistic regression as follows
\begin{equation}\label{item1}
q_{k\ell}=P(x_{k\ell}=1 | y_{k\ell},\theta_k,
R_k=1)=\frac{1}{1+\exp(-(\beta_{\ell 0}+\beta_{\ell
1}\theta_k+\beta_{\ell 2}y_{k\ell}))},
\end{equation}
for  $\ell=1, \dots, m,$ and $k\in r$,
where $\beta_{0\ell}, \beta_{1\ell}$ and $\beta_{2\ell}$ are parameters. Since
$y_{k\ell}$ is known only for units with $x_{k\ell}=1, k\in r$,
Model (\ref{item1}) cannot be estimated. As in the case of unit
nonresponse, we propose to estimate $q_{k\ell}$ as a function of an
auxiliary variable related to the variable of interest, that is
$\theta_k.$ Model (\ref{item1}) is rewritten
\begin{equation}\label{eq:2}
q_{k\ell}=P(x_{k\ell}=1 | \theta_k,
R_k=1)=\frac{1}{1+\exp{(-(\beta_{\ell 0}+\beta_{\ell
1}\theta_{k}))}},
\end{equation}
for  $\ell=1, \dots, m,$ and $k\in r.$ Model (\ref{eq:2}) is not
an ordinary logistic regression model, because the $\theta_k$'s
are unobservable values taken by a latent variable. Latent trait
models can be used in this case to estimate $q_{k\ell},$
$\theta_k$ and the model parameters. Note that in the area of
educational testing and psychological measurement, latent trait
modelling is termed Item Response Theory.

The Rasch model \citep{Rasc:1960} is a first simple latent trait
model that is well known in the psychometrical literature and used
to analyze data from assessments to measure variables such as
abilities and attitudes. It takes the following form
\begin{equation}\label{eq:rasch}
q_{k\ell}=\frac{1}{1+\exp{(-(\beta_{\ell0}+\beta_{ 1}\theta_k))}}
\quad \m{for $\ell=1,\dots,m$ and $k\in r$}.
\end{equation}
The parameters $\beta_{\ell0}$ are estimated for each item $\ell$
and reflect the extremeness (easiness) of item $\ell$: larger
values correspond to a larger probability of a positive response
at all points in the latent space. The parameter $\beta_1$ is
known as the `discrimination' parameter and can be fixed to some
arbitrary value without affecting the likelihood as long as the
scale of the individuals' propensities is allowed to be free. In
many situations the assumption that item discriminations are
constant across items is too restrictive. The two-parameter
logistic (2PL) model generalizes the Rasch model by allowing the
slopes to vary. Specifically, the 2PL model assumes the form given in
equation (\ref{eq:2}).
The parameters $\beta_{\ell1}$ are now estimated for each item
$\ell$ and provide a measure of how much information an item
provides about the latent variable $\theta_k$.  To achieve
identifiability of Model (\ref{eq:2}), we can fix the value of one
or more parameters $\beta_{\ell 0}$ and $\beta_{\ell 1}$ in the
estimation process. \citet{mor:86} showed that in the 2PL model,
all the parameters are identifiable under wide conditions,
provided the number of items exceeds two, and all the slopes are
assumed to be
strictly positive. 
A further generalization to Model (\ref{eq:2}) is considered in
the literature -- the 3PL model -- that includes another
parameter, the \emph{guessing} parameter, to model the probability
that a subject with  a latent variable tending to $-\infty$
responds to an item. Such an extension does not seem necessary in
the context at hand and will not be considered further.

\subsection{Assumptions in latent trait models}

Latent trait models typically rely on the following assumptions. The
first one  is the so-called \emph{conditional independence}
assumption, which postulates that item responses are independent
given the latent variable (i.e. the latent variable accounts for all
association among the observed variables $x_{k\ell}$). Consequently,
given $\theta_k,$ the conditional probability of $\mathbf{x}_k$ is
$$P(\mathbf{x}_k | \theta_k)=\prod_{\ell=1}^m
P(x_{k\ell} | \theta_k).$$
Following \citet[][p. 181]{bar:all:02} `the
assumption of conditional independence can only be tested
indirectly by checking whether the model fits the data. A latent
variable model is accepted as a good fit when the latent variables
account for most of the association among the observed responses.'

A second assumption of Models (\ref{eq:2}) and (\ref{eq:rasch}) is
that of \emph{monotonicity}: as the latent variable $\theta_k$
increases, the probability of response to an item increases or
stays the same across intervals of $\theta_k$. In other words, for
two values of $\theta_k$, say $a$ and $b$, and arbitrarily
assuming that $a< b,$ monotonicity implies that $P(x_{k\ell} = 1|
\theta_k = a)<P(x_{k\ell} = 1| \theta_k = b)$ for $\ell = 1, \dots
, m.$ Larger values of $\theta_k$ are associated with a greater
chance of a response to each item.

Finally, the third, and possibly strongest, assumption of Models
 (\ref{eq:2}) and (\ref{eq:rasch}) is that of
\emph{unidimensionality}, implying that a single latent variable
fully explains the willingness of unit $k$ to answer the
questionnaire. All these basic assumptions imply that the
dependence between the items $x_{k\ell}$ may be explained by the
latent variable $\theta_k$ which represents the units' willingness
and that the probability that a unit $k$ responds to a given
variable increases with $\theta_k$.

\subsection{Estimation of the model}

In what follows we focus on the two-parameter logistic (2PL) model
given in (\ref{eq:2}). Let $\boldsymbol\beta_\ell=(\beta_{\ell 0},
\beta_{\ell 1})'$ and $\boldsymbol\beta=\{\boldsymbol\beta_\ell,
\ell=1, \ldots, m\}.$ Model (\ref{eq:2}) can be fitted using
maximum likelihood or bayesian methods. We focus here on the
former. Under the maximum likelihood approach, three major methods
-- joint, conditional and marginal maximum likelihood -- are
developed. Here, we will concentrate on marginal maximum
likelihood that can be applied to fit the 2PL model. This method
is also used in the simulation studies of  Section \ref{sec:MCsim}.
It consists of maximizing the likelihood of the model after the
$\theta_k$ are integrated out on the basis of a common
distribution assumed on these parameters. In particular, it is
assumed that $\theta_k$ is a random variable following a
distribution with the density function $h(\cdot);$ typically
$\theta_k\sim N(0,1).$ It is also assumed that the response
vectors $\mathbf{x}_k$ are independent of one another and the
conditional independence assumption holds.

For a set of $n_r$ respondents
having the response vectors $\mathbf{x}_k, k = 1, \dots, n_r,$ the
marginal likelihood can be expressed as
$$L(\boldsymbol{\beta}; \mathbf{x}_1, \dots, \mathbf{x}_{n_r}) =\prod_{k=1}^{n_r}
f(\mathbf{x}_k | \boldsymbol{\beta}),$$ where
$f(\mathbf{x}_k | \boldsymbol{\beta})=\int_{-\infty}^\infty
g(\mathbf{x}_k | \theta_k, \boldsymbol\beta) h(\theta_k)
d\theta_k,$ $$g(\mathbf{x}_k | \theta_k,
\boldsymbol\beta)=\prod_{\ell=1}^m
q_{k\ell}^{x_{k\ell}}(1-q_{k\ell})^{1-x_{k\ell}}=\prod_{\ell=1}^m
\frac{\exp\left(x_{k\ell} (\beta_{\ell 0}+\beta_{\ell
1}\theta_k)\right)}{1+\exp(\beta_{\ell 0}+\beta_{\ell
1}\theta_k)},$$ and $h$ now denotes the density of the $N(0,1)$
distribution. The method consists in maximizing the corresponding
log-likelihood, given by
$$\log L(\boldsymbol{\beta}; \mathbf{x}_1, \dots, \mathbf{x}_{n_r})=\sum_{k=1}^{n_r}
\log(f(\mathbf{x}_k | \boldsymbol{\beta})),$$  with respect to
$\boldsymbol{\beta}$ using, for example, the EM algorithm.
Estimates of $\beta_{\ell 0}$ and $\beta_{\ell 1}, \ell=1, \dots,
m$ are thus provided. Afterwards, $\theta_k$ is estimated using
the empirical Bayes method by maximizing the posterior density
$$h(\theta_k | \mathbf{x}_k)=\frac{g(\mathbf{x}_k |
\theta_k, \boldsymbol\beta)h(\theta_k)}{g(\mathbf{x}_k)}\propto
g(\mathbf{x}_k | \theta_k, \boldsymbol\beta)h(\theta_k),$$ with
respect to $\theta_k$ and keeping item parameters and observations
fixed. Estimates of $q_{k\ell}$ are obtained using Expression
(\ref{eq:2}), where $\beta_{\ell 0}, \beta_{\ell 1}$ and
$\theta_k$ are replaced with their estimates.

\subsection{Goodness-of-fit measures of the model}

Different goodness-of-fit measures are proposed in the literature
to test whether the model given in (\ref{eq:2}) adequately fits
the data \citep[see e.g.][]{bar:all:02}. One uses two-way and
three-way margins of the response items. Discrepancies between the
expected ($E$) and observed ($O$) counts in these tables are
measured using the statistic $R=(O-E)^2/{E}$. Large values of $R$
for the second-order or third-order margins will identify sets of
items for which the model does not fit well. Note that the
residuals $(O-E)^2/{E}$ are not independent and they cannot be
summed to give an overall test statistics distributed as a
chi-squared \citep[see][p. 186]{bar:all:02}. Item fit indexes
\citep{bondfox} can be used to this end as well. On the basis of
the estimated latent variables  and item parameters, the expected
response of a unit to an item can be computed. The similarity
between the observed and expected responses to any item can be
assessed  through two fit mean-square statistics: the
outlier-sensitive fit statistic (item outfit) and the
information-weighted fit statistic (item infit). The estimate
produced by the item outfit is relatively more affected by
unexpected responses different from a person's measure, i.e. it is
more sensitive to unexpected observations by units on items that
are relatively very easy or very hard for them to answer. The item
infit has each observation weighted by the information and, on the
other side, is relatively more affected by unexpected responses
closer to a person's measure, i.e. it is more sensitive to
unexpected patterns of observations by units on items that are
roughly targeted on them according to their latent variable value.
The expected value for both statistics is one. For infit and
outfit values greater/less than one indicate more/less variation
between the observed and the predicted response patterns, a range
of 0.5 to 1.5 is generally acceptable \citep{bondfox}.

In addition, point-measure correlations \citep{olsson} can be used
to estimate the correlation between the latent variable and the
single item response. Items for which such measures take  negative
or zero values should be removed from the analysis or may be
evidence that the latent construct is not unidimensional.
Unidimensionality can be tested by running a Principal Components
Analysis of the standardized residuals for the items
\citep{wright96}. In this way the first component (dimension) has
already been removed, and it is possible to look at secondary
dimensions, components or contrasts. Unidimensionality is
supported by observing that the eigenvalue of the first PCA
component in the correlation matrix of the residuals is small
(usually less than 2.0). If not, the loadings on the first
contrast indicate that there are contrasting patterns in the
residuals.

Finally, when items are used to form a scale, they need to have
internal consistency. Cronbach alpha can be used to test whether
items have the reliability property, i.e. if they all measure the
same thing, then they should be correlated with one another.

\subsection{Estimation of $p_k$}\label{sec:esttheta}

Two solution are shown here to estimate $p_k$ using information from the latent trait model.  The first solution
uses logistic regression to estimate $p_k$ for all $k\in s,$ and a
two-stage approach.
\begin{description}
\item[Stage I:] First, an estimate $\widehat{\theta}_k$ of
$\theta_k$ is provided.

To compute a value $\widehat{\theta}_k$ for $k\in \bar r$, we
assume again that unit nonresponse is just an extreme form of item
nonresponse. Thus, a nonrespondent does not answer any item $\ell$
and thus $x_{k\ell}=0,$ for all $\ell=1, \dots, m.$ The
computation of $\widehat{\theta}_k$ for $k\in \bar r$ is handled
as follows: we add to the set $r$ a phantom respondent unit
$\widetilde{k}$ having $x_{\widetilde{k}\ell}$ equal to $0$, for
all $\ell=1, \dots, m.$ We denote this new set by $\widetilde{r}=r
\cup \{\widetilde{k}\}.$ We estimate the parameters of Model
(\ref{eq:2}) using all units $k\in \widetilde{r},$ and compute the
values $\widehat{\theta}_k, k\in \widetilde{r}.$ Model
(\ref{eq:2}) allows the computation of $\widehat{\theta}_k$ for
all $k\in \widetilde{r}.$ Unit $\widetilde{k}$ has an
estimated value $\widehat{\theta}_{\widetilde{k}}.$ We assign to
all units $k\in \bar r$ an estimate $\widehat{\theta}_k$ equal to
$\widehat{\theta}_{\widetilde{k}}.$ Thus, the same value of
$\widehat{\theta}_k$ is provided for all $k\in \bar r$. Using this
method, each unit $k\in s$ has associated an estimate
$\widehat{\theta}_k.$ This is the key feature for the estimation
of the response probabilities
$p_k$ provided in the next stage. 

\item[Stage II:] We use the estimate $\widehat{\theta}_k$, for $k\in s$,
provided in the first stage as a covariate in Model (\ref{nonig22})
instead of the unknown value of $\theta_k$; in particular
\begin{equation}\label{nonig22hat}
p_k=P(R_k=1| \widehat
{\theta}_k)=\frac{1}{1+\exp(-(\alpha_0+\alpha_1\widehat{\theta}_k))},
\mbox{ for all } k\in s.
\end{equation}
Model (\ref{nonig22hat}) provides estimates $\widehat{p}_k$ of
$p_k,$ for all $k\in s.$
\end{description}

A referee suggested the following solution to estimate $p_k.$ Let
$S_k=\sum_{\ell=1}^m x_{k\ell}$ be the raw score for unit $k$,
i.e. the number of items unit $k$ has responded to: if $k\in \bar
r$, then $S_k=0$; if $k\in r,$ then $S_k>0.$ Then $p_{k}$ can be
estimated by modelling $P(S_k>0 | \theta_k).$ By the conditional
independence assumption we have
\begin{eqnarray*}
p_k = P(S_k>0 | \theta_k) & = & 1-P(S_k=0 | \theta_k) =
1-P\left(\cap_{\ell=1}^m (x_{k\ell}=0)\right | \theta_k) \\ & = &
1-\prod_{\ell=1}^m (1-P(x_{k\ell}=1 | \theta_k)).
\end{eqnarray*}
We have $P(x_{k\ell}=1 | \theta_k)=P(R_k=1 | \theta_k)
P(x_{k\ell}=1 | \theta_k, R_k=1)+P(R_k=0 | \theta_k) P(x_{k\ell}=1
| \theta_k, R_k=0) = p_kq_{k\ell},$ because $P(x_{k\ell}=1 |
\theta_k, R_k=0) = 0.$ As a result, we obtain $$p_k =
1-\prod_{\ell=1}^m (1-p_kq_{k\ell}), k\in r.$$

%
\noindent The estimated response probability $\widehat{p}_k, k\in
r$ is obtained as a solution to the polynomial equation
$$\widehat{p}_k=1-\prod_{\ell=1}^m (1-\widehat{p}_k\widehat{q}_{k\ell}). $$
This solution, although very elegant, has two drawbacks. If $m$ is
large, the above polynomial equation is difficult or even
impossible to solve. If it possible to solve the polynomial
equation for moderate $m$, the real solutions are not necessarily
in $(0, 1).$ 
This solution has not been considered here further.

\section{The proposed
estimator and its variance estimation}\label{estimation}

Recall that we have a variable of particular interest $y_j$ and
that item  nonresponse is present for it.  If we
wish to estimate the population total $Y_j$ of
$y_j$, then a naive estimator that does not correct neither for
unit nor for item nonresponse is given by
\begin{equation}\label{naive}
\widehat{Y}_{j,\ml{naive}}=N\sum_{k\in
r_j}\frac{y_{kj}}{\pi_k}/\sum_{k\in r_j} \frac {1}{\pi_k}.
\end{equation}

Reweighting item responders is an also approach to handle item
nonresponse. \citet{mou:kno:00} propose to weight item responders
by the inverse of the fitted probability of item response
$\widehat{q}_{k\ell}$, assuming $\widehat{q}_{k\ell}>0.$
Therefore, a possible adjustment weight for item and unit
nonresponse associated with unit $k\in r_j$ is given by
$1/(\widehat{p}_{k}\widehat{q}_{kj}).$ We propose using the
three-phase estimator adjusted for item and unit nonresponse via
reweighting given by
\begin{equation}\label{htestq}
\widehat{Y}_{j,pq}=\sum_{k\in r_j}
\frac{y_{kj}}{\pi_k\widehat{p}_k\widehat{q}_{kj}},
\end{equation}
 where $\widehat{p}_k$ is
provided by Model (\ref{nonig22hat}), and $\widehat{q}_{kj}$ by
Model (\ref{eq:2}). Proposals that use imputation of $y_{kj}$
values for $k\in r\backslash r_j$ to deal with item nonresponse
are also considered but not reported for reasons of space. They
are available from the authors upon request.

The properties of the proposed estimator (\ref{htestq}) depend on
the assumptions made about the unit and the item nonresponse
mechanisms. In particular, Estimator (\ref{htestq}) assumes a
second phase of sampling with unknown response probabilities. If
we ignore estimation of $\theta_k$ in Model (\ref{nonig22hat}),
the results in \citet{kim:kim:07} on design consistency of the
two-phase estimator that uses estimated response probabilities
hold here as well when considering  maximum likelihood estimates
for the parameters $\alpha_{ 0}$ and $\alpha_1$. Again, ignoring
estimation of the latent variable $\theta_k$ and using marginal
maximum likelihood estimates for the parameters $\beta_{\ell 0}$
and $\beta_{\ell 1}$ in Model (\ref{eq:2}),  estimator
$\widehat{Y}_{j,pq}$ will be consistent if the models for unit and
item nonresponse probabilities are correctly specified.

We can consider replication methods for variance estimation of the
proposed estimator and combine proposals for two-phase sampling
\citep{kim2006replication} and for generalized calibration in the
presence of nonresponse \citep{kot:06}. In particular, the
replicate variance estimator can be written as
$$\widehat{V}_r=\sum_{l=1}^Lc_l\Big(\widehat{Y}_{j,pq}^{(l)}-\widehat{Y}_{j,pq}\Big)^2,$$
where $\widehat{Y}_{j,pq}^{(l)}$ is the $l$-th version of
$\widehat{Y}_{j,pq}$ based on the observations included in the
$l$-th replicate, $L$ is the number of replications, $c_l$ is a
factor associated with replicate $l$ determined by the replication
method. The $l$-th replicate of $\widehat{Y}_{j,pq}$ can be
written as $\widehat{Y}_{j,pq}^{(l)}=\sum_{k\in r_j}w^{(l)}_{3k}
y_{kj}$, where $w^{(l)}_{3k}$ denotes the replicate weight for the
$k$-th unit in the $l$-th replication. These replicate weights are
computed using a two-step procedure. 

First, note that, if we
ignore for the moment the presence of item nonresponse, the
two-phase estimator $\widehat{Y}_{j,p}=\sum_{k\in r}
w_{2k}y_{kj}$, has weights
$$w_{2k}=1/(\pi_kp_k)=w_{1k}F(\widehat{\theta}_k;\alpha_0,\alpha_1),$$
with $w_{1k}=1/\pi_k$,
$F(\widehat{\theta}_k;\alpha_0,\alpha_1)=1+\exp(-(\alpha_0+\alpha_1\widehat{\theta}_k))$
(see Equation (\ref{nonig22hat})). Let $\ha{z}_1=\sum_{k\in
s}w_{1k}\b{z}_{1k}$ be the first phase estimate of the total of
variable $\b z_1$ defined as
$\b{z}_{1k}=\pi_kp_k(1,\widehat{\theta}_k)^T$. Then, parameters
$\alpha_0$ and $\alpha_1$ are such that
\begin{equation}\label{eq:jackb1}
\sum_{k\in
r}w_{1k}F(\widehat{\theta}_k;\alpha_0,\alpha_1)\b{z}_{1k}=\ha{z}_1.
\end{equation}
This procedure is equivalent to obtaining unweighted maximum likelihood estimates, but is convenient to set it as a non-linear generalized calibration problem. In this way, it is possible to use the approach in \citet{kot:06}, combined with that in \citet{kim2006replication}, to obtain replicate weights using the following steps.
\begin{description}
\item[Step 1.] Compute the first phase estimate of the total of
$\b{z}_{1k}$ with $l$-th observation deleted, i.e.
$\ha{z}_1^{(l)}=\sum_{k\in s}w_{1k}^{(l)}\b{z}_{1k}$, where
$w_{1k}^{(l)}$  is the classical jackknife replication weight for
unit $k$ in replication $l$. Compute the jackknife weights for the
second phase sampling using $\ha{z}_1^{(l)}$ as a benchmark. In
particular, $w_{2k}^{(l)}$ are chosen to be
$w_{2k}^{(l)}=w_{2k}w_{1k}^{(l)}F(\widehat{\theta}_k;\alpha_0,\alpha_1)/w_{1k}$
with $\alpha_0$ and $\alpha_1$ such that $$\sum_{k\in
r}w_{2k}^{(l)}\b{z}_{1k}=\ha{z}_1^{(l)}.$$ This procedure provides
weights that are very similar to those considered in
\citet{kot:06} and can be computed using existing software that
handles generalized calibration.
\end{description}
Item nonresponse is handled similarly by considering
$w_{3k}=1/(\pi_kp_kq_{kj})=w_{2k}F(\widehat{\theta}_k; \beta_{j0},\beta_{j1})$ (compare Equation (\ref{eq:rasch})). A
major approximation here is to assume that, given
$\widehat{\theta}_k$, parameters $\beta_{j0}$ and $\beta_{j1}$ are
estimated using a classical logistic model (instead of a 2PL
model) and are such that $$\sum_{k\in
r_j}w_{2k}F(\widehat{\theta}_k; \beta_{j0},\beta_{j1})
\b{z}_{2k}=\ha{z}_2,$$ where $\ha{z}_2=\sum_{k\in
r}w_{2k}\b{z}_{2k}$ and
$\b{z}_{2k}=\pi_kp_kq_{kj}(1,\widehat{\theta}_k)^T$. Another
drawback is that auxiliary variables $\b{z}_{2k}$ depend on $j$
and, therefore, different sets of weights have to be produced for
the different variables of interest.

\begin{description}

\item[Step 2.] Third phase jackknife weights are obtained by first
computing the second phase estimate of the total of $\b{z}_{2k}$
with unit $l$ removed by using weights coming from Step 1, i.e.
$\ha{z}_2^{(l)}=\sum_{k\in r}w_{2k}^{(l)}\b{z}_{2k}$. Then, using
$\ha{z}_2^{(l)}$ as a benchmark, $w_{3k}^{(l)}$ are chosen to be
$w_{3k}^{(l)}=w_{3k}w_{2k}^{(l)}F(\widehat{\theta}_k;\beta_{j0},\beta_{j1})/w_{2k}$
with $\beta_{j0}$ and $\beta_{j1}$ computed via $$\sum_{k\in
r_j}w_{3k}^{(l)}\b{z}_{2k}=\ha{z}_2^{(l)}. $$
\end{description}



\section{Simulation studies}\label{sec:MCsim}
We evaluate the performance of the estimator presented in Section
\ref{estimation} by means of A Monte Carlo simulation under two
different settings. The first one uses a real data set as the
population and considers variables of interest that are all
binary, while the second one uses simulated population data with
variables of interest that are  continuous. Results from the
first setting are presented in Section \ref{sec:abort}, while
those from the second setting are presented in Section
\ref{sec:setting2}.

In both settings, simple random sampling without replacement is
employed  and the following estimators are considered:
\begin{itemize}
\item $HT=\sum_{k\in s} y_{kj}/\pi_k$: the Horvitz-Thompson
estimator in the case of full response is computed as a benchmark
in the absence of nonresponse. \item $\widehat{Y}_{j,\ml{naive}}$:
the naive estimator given in (\ref{naive}); no explicit action is
taken to adjust for unit and item nonresponse. Note that for
simple random sampling without replacement, it reduces to
$\widehat{Y}_{j,\ml{naive}}=N\sum_{k\in r_j} y_{kj}/n_{r_j},$
where $n_{r_j}$ is the size of the set $r_j,$ and it is the same
as the Horvitz-Thompson estimator adjusted for unit nonresponse
that assumes uniform response probabilities estimated by
$n_{r_j}/n.$ \item $\widehat{Y}_{j,pq}$: the three-phase estimator
proposed in Section \ref{estimation}, Equation (\ref{htestq}).
\item $\widehat{Y}_{j,pq,\ml{true}}$: the three-phase estimator
that uses the true values for the response probabilities $p_k$ and
$q_{kj}$ is also computed for comparison with $\widehat{Y}_{j,pq}$
to understand the effect of estimating the response probabilities.
\end{itemize}

The simulations are carried out in R version 2.15, using the
R package \texttt{ltm} \citep{riz:06} to fit the latent trait models. The following performance measures are computed for each estimator,
generically denoted below by $\widehat{Y}$ where suffix $j$ is dropped
for ease of notation ($Y$ denotes the population total):
\begin{itemize}
\item the Monte Carlo Bias
$$\m{B}=E_{sim}(\widehat{Y})-Y,$$ where $E_{sim}(\widehat{Y})=\sum_{i=1}^{M} \widehat{Y}_i/M,$   $\widehat{Y}_i$ is the value
of the estimator $\widehat{Y}$ at the $i$-th simulation run and $M$ is total number of simulation runs;
\item the Relative Bias $$\m{RB}=\frac{\m{B}}{Y};$$
\item the
Monte Carlo Standard Deviation
$$\sqrt{\m{VAR}}=\sqrt{\frac{1}{M-1}\sum_{i=1}^{M}\left(\widehat{Y}_i - E_{sim}(\widehat{Y})\right)^2};$$ \item
the Monte Carlo Mean Squared Error
$$\m{MSE}=\m{B}^2+\m{VAR}.$$
\end{itemize}

\subsection{Simulation setting 1} \label{sec:abort}
We consider the Abortion data set formed by four binary variables
extracted from the 1986 British Social Attitudes Survey and
concerning the attitude towards abortion. The data is available in
the R package `ltm' \citep{riz:06}. $N=379$ individuals answered
the following questions after being asked if the law should allow
abortion under the circumstances presented under each item:
\begin{enumerate}
\item The woman decides on her own that she does not wish to keep the baby.
\item The
couple agree that they do not wish to have a child. \item The
woman is not married and does not wish to marry the man. \item The
couple cannot afford any more children.
\end{enumerate}
The variable of interest $y_{j}$ is selected to be  the second one ($j=2$) with a total  $Y_j=225$ in the population.

The data is analyzed by \citet{bar:all:02} as an example in which
a latent variable can be found that measures the attitude towards
abortion. At the population level, we  compute the latent variable
(denoted here by $\theta^a_k$) using Model (\ref{eq:2}) on the
$\{y_{k\ell}\}_{k=1,\ldots,N; \ell=1,\ldots, 4}$ data. The
correlation between the values $y_{k\ell}$ and $\theta^a_k$ is
approximatively equal to 0.85, for  $\ell=1, \dots, 4.$
Afterwards, we have set $\theta_k=\widehat{\theta}^a_k,$ for all
$k=1,\ldots,N.$

At the population level, the unit response probabilities are
generated using the following response model
\begin{equation}\label{abortion}
p_k=1/(1+\exp(-(0.7+y_{k2}+\theta_k+0.2\varepsilon_k))),
\end{equation}
with $\varepsilon_k\sim U(0,1)$, to simulate nonignorable
nonresponse. The population mean of $p_k$ is approximately 0.74.

To generate item response probabilities at the population level,
the following model is used
\begin{equation}\label{abortion2}
q_{k\ell}=1/(1 +\exp(-(b_\ell\theta_k+a_\ell+y_{k\ell}))), \quad
\m{for }\ell=1,\ldots,4,
\end{equation}
where $b_\ell=3$, for $\ell=1,\dots, 4$, while $a_\ell$ takes different values according to $\ell$; in particular, $a_1=1$, $a_2=0$, $a_3=-0.5$ and $a_4=1$. The nominal  item
nonresponse rate for the four items in the population is 35\%,
42\%, 47\%, 31\%, respectively.

We draw $M=10,000$ simple random samples without replacement
from the population using two sample sizes: $n=50$
and $n=100.$ In each sample $s,$ the units are classified as
respondents according to Poisson sampling, using the probabilities
$p_k$ computed as in Equation (\ref{abortion}) and resulting in
the set $r$. Then, given $r$, the matrix $\{x_{k\ell}\}_{k\in r;
\ell=1,...,4}$ is constructed where the values $x_{k\ell}$ are
drawn using Poisson sampling with probabilities $q_{k\ell}$
defined in (\ref{abortion2}). In each simulation run, Model
(\ref{eq:2}) and the respondents set $r$ are used to compute the
variable $\widehat{\theta}_k$ for all $k\in s$ as described in
Section \ref{sec:esttheta}. Model (\ref{nonig22hat}) is fitted to
obtain $\widehat{p}_k.$ The average item nonresponse rate over
simulations for the four items is found to be 26\%, 33\%, 38\% and
23\%. The jackknife variance estimator was computed as described
in Section \ref{estimation} using the \texttt{gencalib()} function in R
package `sampling' \citep{til:mat:12} and the logistic distance
\citep{dev:sar:sau:93}.

Table \ref{tab:setting1} reports the results for $n=50$ and
$n=100$. As expected, $HT$ and $\widehat{Y}_{j,pq,\ml{true}}$ have
almost zero bias, with the second one shows a relatively larger
MSE that is due uniquely to the smaller sample size. The naive
estimator shows a very large negative bias. This is due to the fact that
units with a zero value of $y_j$ are less likely to respond and
the total is clearly underestimated. The estimator
$\widehat{Y}_{j, pq}$ shows a much smaller bias than the naive
estimator. 
Note that the performance of the proposed estimator is mostly
driven by absolute bias, so that the performance is not
particularly different when increasing the sample size, apart from
a decrease in variance. If we compare
$\widehat{Y}_{j,pq,\ml{true}}$ and $\widehat{Y}_{j,pq}$, 
we note that $\widehat{Y}_{j,pq}$ still suffers from some bias
that comes from response model misspecification (we are not
accounting for the variables of interest values). 

For the proposed estimator, the jackknife variance estimator was also tested by looking at the empirical coverage of a 95\% confidence interval computed for each replicate as $\widehat{Y}_{j,pq}\pm1.96 \sqrt{\hat V_r}$. For $n=50,$ the
mean value of $\sqrt{\hat{V}_{r}}$ over simulations was 54.8,
while for $n=100,$ 53.3, with a 95\% coverage rate of 94.6\% and
96.3\%, respectively. The replicate estimator overestimates the
Monte Carlo standard deviation reported for $\widehat{Y}_{j,pq}$
in Table \ref{tab:setting1} in both cases, but shows  good
coverage rates.

\begin{table}[htb!]
\caption{Simulation results for setting 1 -- Abortion data set }\label{tab:setting1}
\begin{center}
\begin{tabular}{lrrrr}
  Estimator         & B  &    $\sqrt{\m{VAR}}$ & MSE &  \% RB   \\ \hline
$n=50$ \\
  $HT$                                   &  0.05     & 24.5   & 600.5  & $<0.1$   \\
  $\widehat{Y}_{j,\ml{naive}}$           & -126.5    & 19.4   & 16378.6& -56.2  \\
  $\widehat{Y}_{j,pq}$                   &  20.6     & 32.4   & 1474.1 & 9.1    \\
  $\widehat{Y}_{j,pq,\ml{true}}$         &  0.02     & 35.0   & 1225.0 & $<0.1$   \\
$n=100$ \\

  $HT$                            &   -0.06 & 16.0  & 255.5   & $<0.1$ \\
  $\widehat{Y}_{j,\ml{naive}}$    & -126.9  & 13.5  &16284.1  & -56.4 \\
  $\widehat{Y}_{j,pq}$            &  17.9   & 21.9  &802.2    &8.0   \\
  $\widehat{Y}_{j,pq,\ml{true}}$  &   -0.1  & 23.7  & 559.9   &$<0.1$   \\ \hline
\end{tabular}
\end{center}
\end{table}

To study the performance of the latent model on the population
level and the correlation between the variable of interest and the
estimated latent variable, we  apply  the procedure described
earlier using $q_{k\ell}$ defined in (\ref{abortion2}) to
construct the matrix $\{x_{k\ell}\}_{k=1,\ldots,N; \ell=1,...,4}$
for all population units. We fit Model (\ref{eq:2}) on the
population level and compute the variable $\theta_k$ for all
$k=1,\ldots,N$.  The Cronbach's alpha measure takes value 0.83
showing a good internal consistency of the items.
The correlation coefficient between the variable of interest and
the estimated latent variable takes value 0.76, indicating that
the latent auxiliary information has a strong power of predicting
$y_{k2},$ as advocated in the model of \citet{cas:sar:wre:83}.
Inspection of the two-way margins for the matrix $\{x_{k\ell}\}$
gives the residuals $(O-E)^2/E$ between 0.03 and 0.23. Similarly,
the three-way margins for the matrix $\{x_{k\ell}\}$ give
residuals between 0 and 1.19. This indicates that we have no
reason to reject here the one-factor latent Model
(\ref{eq:2}) 
\citep[see][p. 186]{bar:all:02}.

\subsection{Simulation setting 2} \label{sec:setting2}
We generate $\{y_{k1},\ldots,y_{k6},\theta_k\}$ for
$k=1,\ldots,N=2,000$ using a multivariate normal distribution with
mean 1. The degree of correlation between $y_{\ell}$ and
$y_{\ell'}$ is 0.8, with $\ell, \ell'=1,\ldots,6, \ell\neq\ell'$.
We set the variable of interest to be $y_6$ and consider different
degrees of correlation between $y_{6}$ and
$\boldsymbol\theta=(\theta_k)'$ namely, 0.3, 0.5, 0.8. The values
of $\theta_k$ are afterwards standardized to have mean 0 and
variance 1.

The response probabilities are obtained by first computing
\begin{equation}\label{sim2p}
p^\circ_k=1/(1+\exp(-(0.5+y_{k1}+\theta_k))), \quad \m{for $k=1,\ldots,N$},
\end{equation}
and then rescaling them to take values between 0.1 and 0.9 using the transformation
\begin{equation}\label{sim2p1}
p_k=(p^\circ_k-\min_k p^\circ_k)/(\max_k
p^\circ_k-\min_k p^\circ_k)\times0.8+0.1,
\end{equation}
with a population mean approximatively equal to 0.7.

The item response probabilities are generated by first computing
\begin{equation}\label{latent2}
q^\circ_{k\ell}=1/(1 +\exp(-(b_\ell\theta_k+a_\ell+y_{k\ell}))),\quad \m{for $k=1,\ldots,N$ and $\ell=1,\ldots,6$,}
\end{equation}
where $\{a_\ell\}_{\ell=1,\dots, 6}=\{1, 0, -0.5, 1, 0,
-0.5\}$ and $\{b_\ell\}_{\ell=1,\dots, 6}=\{1, 1, 1, 1.5,
1.5, 1.5\},$ and then rescaling the values to be between 0.1
and 0.95 using the transformation
\begin{equation}\label{latent21}
q_{k\ell}=(q^\circ_{k\ell}-\min_k q^\circ_{k\ell})/(\max_k
q^\circ_{k\ell}-\min_{k}q^\circ_{k\ell})\times0.85+0.1.
\end{equation}

We draw $M=10,000$ samples by simple random sampling without
replacement of size $n=200$. For each sample $s$, a response set
$r$ is created by carrying out Poisson sampling with parameter
${p}_{k}$ defined in (\ref{sim2p}). Each element of the matrix
$\{x_{k\ell}\}_{k\in r,\ell=1,\ldots,6}$ is  generated using
Poisson sampling with parameter ${q}_{k\ell}$ defined in
(\ref{latent2}). Item nonresponse rates over simulations take
approximately value 18\%, 28\%, 35\%, 19\%, 29\%, 34\%, for
$\ell=1,\ldots,6$, respectively. For  each simulation run, Model
(\ref{eq:2}) is used to compute the variable $\widehat{\theta}_k$
for all $k\in s$. Model (\ref{nonig22hat}) is then  fitted to
obtain $\widehat{p}_k.$

\begin{table}[t]
\begin{center}
\caption{Simulation results for setting 2 -- Simulated continuous data}\label{tab11}
\begin{tabular}{lrrrr}
Estimator         & B  &    $\sqrt{\m{VAR}}$ & MSE &  RB\%  \\
\hline
correlation coefficient 0.3\\
 $HT$                                 &   -0.7   &  131.6  &17331.2  & $<-0.0$       \\
 $\widehat{Y}_{j,\ml{naive}}$         &  825.6   &  177.1  &713039.3   & 41.0       \\
 $\widehat{Y}_{j,pq}$                 & -227.4   &  188.0  &87033.0    & -11.3     \\
 $\widehat{Y}_{j,pq,\ml{true}}$       &   48.4   &  231.8  &56073.2    &$2.4$      \\

correlation coefficient 0.5\\
 $HT$                             & $0.1$   & 135.0&18220.5 &$0.0$   \\
 $\widehat{Y}_{j,\ml{naive}}$     & 972.6  & 176.2&977009.5&50.7        \\
 $\widehat{Y}_{j,pq}$             & -180.0 & 175.5&63552.0&$-9.4$ \\
 $\widehat{Y}_{j,pq,\ml{true}}$   & 74.8   & 212.7&50844.0 &$3.9$ \\

correlation coefficient 0.8 \\
  $HT$                                &$-0.1$ & 134.1&17992.0 &$<-0.0$\\
 $\widehat{Y}_{j,\ml{naive}}$         &1154.6 & 168.1&1361388.1&57.7     \\
 $\widehat{Y}_{j,pq}$                 &-184.8 & 164.4&61173.0&-9.2 \\
 $\widehat{Y}_{j,pq,\ml{true}}$       &100.6  & 196.2&48597.9 &5.0\\
\hline
\end{tabular}
\end{center}
\end{table}

Table \ref{tab11} reports on the performance of the estimators for
the three values taken by the nominal correlation coefficient
between $y_{k1}$ and $\theta_k:$ 0.3, 0.5, and 0.8. The proposed
estimator is always able to reduce bias over the naive estimator,
even when the correlation between the variable of interest and the
latent variable gets smaller. The relative bias takes acceptable
values in most cases. Bias deserves a closer look. The naive estimator in all cases
largely overestimates the total. This is expected, because the
values $p_k$, $q_{k6}$, $\theta_k$ and $y_{k6}$ all go in the same
direction. Therefore, in our respondents sample, we are more
likely to find relative larger values for $y_6$ by this providing
overestimation for the naive estimator. On the other hand, $\widehat{Y}_{j,pq}$ underestimates the total
because it is based only on the observed units of  $r_j$ that do
have relatively large values for $y_6$, but also relatively large
values for $p_k$ and $q_{k6}$ and, therefore, end up having a
small weight.

The matrix of population values $\{x_{k\ell}\}_{k=1,\ldots, 2000,
\ell=1,\dots, 6}$ is constructed in the same way as in Section
\ref{sec:abort} to validate the assumptions behind the 2PL model.
The Cronbach's alpha takes approximately value 0.5 for the
correlation coefficient equal to 0.3, 0.6 for 0.5, and 0.7 for
0.8; the pairwise association between the six items reveals
p-values smaller than $0.01.$ Inspection of the two-way and
three-way margins of the matrix $\{x_{k\ell}\}$ gives residuals
$(O-E)^2/E$ that all take values smaller than 4. Therefore, the
one factor latent model can be accepted and items all seem to be
measuring the same latent trait.

\section{Discussion and conclusions}\label{sec:concl}
We have proposed a reweighting system to compensate for
non-ignorable nonresponse based on a latent auxiliary variable.
This variable is computed for each unit in the sample using a
latent model assuming the existence of item nonresponse and that
the same latent structure is hidden behind item and unit
nonresponse. Unit response probabilities are then estimated by a
logistic model that uses as a covariate the latent trait extracted
by the response patterns using a latent trait model. The proposed
reweighting system is then  used in a three-phase estimator to
handle nonresponse, together with a replication method to estimate its uncertainty. The main goal is to reduce the nonresponse
bias in the estimation of the population total. The proposed
estimator performs well in our simulation studies compared with
the naive estimator, and the gain in efficiency is substantial in
certain cases. Reductions in bias are also seen when the
correlation between the latent trait and the variable of interest
is modest.


By design, the estimated latent variable $\widehat{\theta}_k$ is
related to the response indicators $x_{kj}$ for the variable of
interest $y_j$; since  nonresponse is assumed to be non-ignorable,
$y_{kj}$ and $x_{kj}$ are related as well. If the following
condition holds,
$$\rho_{{y}_j,{x}_j}^2+\rho_{\widehat\theta, x_j}^2>1,$$ where the correlation
coefficients $\rho_{{y}_j,{x}_j},\rho_{\widehat\theta, x_j}>0$,
then ${y}_j$ and $\widehat\theta$ are positively correlated
\citep[see][]{lan:sch:owe:01}. Note that the minimum degree of
correlation between the variable of interest and the latent variable 
capable of reducing the nonresponse bias was found to be 0.3 in
the simulation setting 2.

We have considered the case in which no auxiliary information is
available at the sample or population level to reduce nonresponse
bias. Observed covariates (if available) and the latent variable
 can be, however, used together in the
estimation of response probabilities. Moreover, latent trait
models can, themselves,  be fitted with covariates. The introduction of
covariates in these models should be carried out with increasing
prudence of variance.

The proposed estimator is a three-phase estimator using a
reweighting system based on $\widehat{p}_k$ and
$\widehat{q}_{kj}.$ It is known that small values of
$\widehat{p}_k$ and $\widehat{q}_{kj}$ may lead to unstable
reweighted estimators because of large nonresponse weights.  To
overcome this problem, the propensity score method
\citep[e.g.][]{elt:yan:97} is often used in practice, providing a
good solution against extreme weights adjustments. In order to
apply this method in our framework, the respondents to $y_j$
should be grouped in different classes given by the quantiles of
$1/(\widehat{p}_k\widehat{q}_{kj}).$ The final step is the
calculation of a weight for each class.

Final remarks concern the conditional independence assumption in
latent trait models. In nonresponse literature, it is usual to use
Poisson sampling to model unit response behavior by assuming that
units in the set $r$ are selected with unknown response
probabilities and that response is independent from unit to unit.
The conditional independence assumption in the latent trait models
is a similar condition applied to items. Both assumptions are
strong, sometimes they are in doubt, yet they are necessary in the
statistical inferential process.

Different methods were developed in psychometric literature to
relax the conditional independence assumption. We cite here the
\emph{partial independence} approach by \citet{rea:rau:06},
developed for the case where responses to earlier questions
determine whether later questions are asked or not, and where the
usual conditional independence assumption of standard models
fails. This approach could be used in our framework for the case
where $q_{k\ell}$ is defined as $P(x_{k\ell}=1 | x_{kj}, \mbox{
for some }  j\in\{1, \dots, m\}, \ell\neq j, \theta_k)$ instead of
$P(x_{k\ell}=1 | \theta_k), k\in r.$ Another useful approach for
cases where items are clustered is the latent trait hierarchical
modeling. A random effect is introduced into a latent trait model to
account for potential residual dependence due to the common
sources of variation shared by clusters of items \citep[see
e.g.][]{sco:ip:02}. Further research should be done to accommodate
these approaches in the survey sampling framework.

\bibliographystyle{apalike}
\bibliography{bibyves,bibyves2}

\begin{thebibliography}{}

\bibitem[Bartholomew et~al., 2002]{bar:all:02}
Bartholomew, D.~J., Steele, F., Moustaki, I., and Galbraith, J.~I. (2002).
\newblock {\em The Analysis and Interpretation of Multivariate Data for Social
  Scientists}.
\newblock Chapman and Hall/CRC.

\bibitem[Beaumont, 2000]{beu:00}
Beaumont, J.~F. (2000).
\newblock An estimation method for nonignorable nonresponse.
\newblock {\em Survey Methodology}, 26:131--136.

\bibitem[Bethlehem, 1988]{bet:88}
Bethlehem, J. (1988).
\newblock Reduction of nonresponse bias through regression estimation.
\newblock {\em Journal of Official Statistics}, 4(3):251--260.

\bibitem[Biemer and Link, 2007]{bie:lin:07}
Biemer, P.~P. and Link, M.~W. (2007).
\newblock {\em Evaluating and modeling early cooperator effects in RDD
  surveys}.
\newblock New York: Wiley.

\bibitem[Bond and Fox, 2007]{bondfox}
Bond, T. and Fox, C. (2007).
\newblock {\em Applying the Rasch model: Fundamental measurement in the human
  sciences (2nd ed.)}.
\newblock Lawrence Erlbaum Associates, Inc, Mahwah, NJ.

\bibitem[Cassel et~al., 1983]{cas:sar:wre:83}
Cassel, C.~M., S\"arndal, C.~E., and Wretman, J.~H. (1983).
\newblock Some uses of statistical models in connection with the nonresponse
  problem.
\newblock In Madow, W.~G. and Olkin, I., editors, {\em Incomplete Data in
  Sample Surveys}, volume~3, pages 143--160. New York: Academic Press.

\bibitem[Chambers and Skinner, 2003]{cha:ski:03}
Chambers, R.~L. and Skinner, C. (2003).
\newblock {\em Analysis of Survey Data}.
\newblock Wiley, New York.

\bibitem[Copas and Farewell, 1998]{cop:far:98}
Copas, A.~J. and Farewell, V.~T. (1998).
\newblock Dealing with non-ignorable non-response by using an
  'enthusiasm-to-respond' variable.
\newblock {\em Journal Of The Royal Statistical Society, Series A},
  161:385--396.

\bibitem[De~Menezes and Bartholomew, 1996]{dem:bar:96}
De~Menezes, L.~M. and Bartholomew, D.~J. (1996).
\newblock New developments in latent structure analysis applied to social
  attitudes.
\newblock {\em Journal of Royal Statistical Society A}, 159:213--224.

\bibitem[Deville et~al., 1993]{dev:sar:sau:93}
Deville, J.-C., Sï¿½rndal, C.-E., and Sautory, O. (1993).
\newblock Generalized raking procedure in survey sampling.
\newblock {\em Journal of the American Statistical Association}, 88:1013--1020.

\bibitem[Drew and Fuller, 1980]{dre:ful:80}
Drew, J.~H. and Fuller, W.~A. (1980).
\newblock Modeling nonresponse in surveys with callbacks.
\newblock {\em Proceedings of the Section on Survey Research Methods of the
  American Statistical Association}.

\bibitem[Eltinge and Yansaneh, 1997]{elt:yan:97}
Eltinge, J.~L. and Yansaneh, I.~S. (1997).
\newblock Diagnostics for formation of nonresponse adjustment cells, with an
  application to income nonresponse in the u. s. consumer expenditure survey.
\newblock {\em Survey Methodology}, 23:33--40.

\bibitem[Greenlees et~al., 1982]{gre:ree:zie:82}
Greenlees, J.~S., Reece, W.~S., and Zieschang, K.~D. (1982).
\newblock Imputation of missing values when the probability of response depends
  on the variable being imputed.
\newblock {\em Journal of the American Statistical Association}, 77:251--261.

\bibitem[Groves, 2006]{gro:06}
Groves, R.~M. (2006).
\newblock Nonresponse rates and nonresponse bias in household surveys.
\newblock {\em Public Opinion Quarterly}, 70 (5):646--675.

\bibitem[Groves et~al., 2006]{gro:all:06}
Groves, R.~M., Couper, M., Presser, S., Singer, E., Tourangeau, R., Acosta,
  G.~P., and Nelson, L. (2006).
\newblock Experiments in producing nonresponse bias.
\newblock {\em Public Opinion Quarterly}, 70(5):720--736.

\bibitem[Kim and Kim, 2007]{kim:kim:07}
Kim, J.~K. and Kim, J.~J. (2007).
\newblock Nonresponse weighting adjustment using estimated response
  probability.
\newblock {\em Canadian Journal of Statistics}, 35:501--514.

\bibitem[Kim et~al., 2006]{kim2006replication}
Kim, J.~K., Navarro, A., and Fuller, W.~A. (2006).
\newblock Replication variance estimation for two-phase stratified sampling.
\newblock {\em Journal of the American Statistical Association},
  101(473):312--320.

\bibitem[Kott, 2006]{kot:06}
Kott, P.~S. (2006).
\newblock Using calibration weighting to adjust for nonresponse and coverage
  errors.
\newblock {\em Survey Methodology}, 32:133--142.

\bibitem[Langford et~al., 2001]{lan:sch:owe:01}
Langford, E., Schwertman, N., and Owens, M. (2001).
\newblock Is the property of being positively correlated transitive?
\newblock {\em The American Statistician}, 55 (4):322--325.

\bibitem[Legg and Fuller, 2009]{legg2009two}
Legg, J.~C. and Fuller, W.~A. (2009).
\newblock Two-phase sampling.
\newblock {\em Handbook of statistics}, 29:55--70.

\bibitem[Little and Vartivarian, 2005]{lit:var:05}
Little, R. and Vartivarian, S. (2005).
\newblock Does weighting for nonresponse increase the variance of survey means?
\newblock {\em Survey Methodology}, 31:161--168.

\bibitem[Little and Rubin, 1987]{lit:rub:87}
Little, R. J.~A. and Rubin, D.~B. (1987).
\newblock {\em Statistical Analysis with Missing Data}.
\newblock John Wiley \& Sons, New York.

\bibitem[Moran, 1986]{mor:86}
Moran, P. A.~P. (1986).
\newblock Identification problems in latent trait models.
\newblock {\em British Journal of Mathematical and Statistical Psychology}, 39,
  Issue 2:208--212.

\bibitem[Moustaki and Knott, 2000]{mou:kno:00}
Moustaki, I. and Knott, M. (2000).
\newblock Weighting for item non-response in attitude scales using latent
  variable models with covariates.
\newblock {\em Journal of Royal Statistical Society, Series A}, 163:445--459.

\bibitem[Oh and Scheuren, 1983]{oh:sch:83}
Oh, H.~L. and Scheuren, F.~J. (1983).
\newblock Weighting adjustments for unit non-response.
\newblock In Madow, W.~G., Olkin, I., and Rubin, D.~B., editors, {\em
  Incomplete Data in Sample Surveys}, volume~2, pages 143--184. New York:
  Academic Press.

\bibitem[Olsson et~al., 1982]{olsson}
Olsson, U., Drasgow, F., and Dorans, N. (1982).
\newblock The {P}olyserial {C}orrelation {C}oefficient.
\newblock {\em Psychometrika}, 47:337--347.

\bibitem[Qin et~al., 2002]{qin:leu:sha:02}
Qin, J., Leung, D., and Shao, J. (2002).
\newblock Estimation with survey data under nonignorable nonresponse or
  informative sampling.
\newblock {\em Journal of the American Statistical Association}, 97:193--200.

\bibitem[Rasch, 1960]{Rasc:1960}
Rasch, G. (1960).
\newblock {\em Probabilistic Models for Some Intelligence and Attainment
  Tests}.
\newblock The Danish Institute of Educational Research, Copenhagen.

\bibitem[Reardon and Raudenbush, 2006]{rea:rau:06}
Reardon, S.~F. and Raudenbush, S.~W. (2006).
\newblock A partial independence item response model for surveys with filter
  questions.
\newblock {\em Sociological Methodology}, 36, no. 1:257--300.

\bibitem[Rizopoulos, 2006]{riz:06}
Rizopoulos, D. (2006).
\newblock ltm: An {R} package for latent variable modelling and item response
  theory analyses.
\newblock {\em Journal of Statistical Software}, 17 (5):1--25.

\bibitem[Rosenbaum and Rubin, 1983]{ros:rub:83}
Rosenbaum, P.~R. and Rubin, D.~B. (1983).
\newblock The central role of the propensity score in observational studies for
  causal effects.
\newblock {\em Biometrika}, 70:41--55.

\bibitem[S\"{a}rndal and Lundstr\"{o}m, 2005]{sar:lun:05}
S\"{a}rndal, C.~E. and Lundstr\"{o}m, S. (2005).
\newblock {\em Estimation in Surveys with Nonresponse}.
\newblock John Wiley \& Sons, New York.

\bibitem[Scott and Ip, 2002]{sco:ip:02}
Scott, S.~L. and Ip, E.~H. (2002).
\newblock Empirical bayes and item-clustering effects in a latent variable
  hierarchical model: A case study from the national assessment of educational
  progress.
\newblock {\em Journal of American Statistical Association}, 97, no. 459:1--11.

\bibitem[Skrondal and Rabe-Hesketh, 2007]{skr:rab:07}
Skrondal, A. and Rabe-Hesketh, S. (2007).
\newblock Latent variable modelling: A survey.
\newblock {\em Scandinavian Journal of Statistics}, 34:712--745.

\bibitem[Tillé and Matei, 2012]{til:mat:12}
Tillé, Y. and Matei, A. (2012).
\newblock {\em sampling: Survey Sampling}.
\newblock R package version 2.5.

\bibitem[Wright, 1996]{wright96}
Wright, B. (1996).
\newblock Local dependency, correlations and principal components.
\newblock {\em Rasch Meas Trans}, 10-3:509--511.

\bibitem[Zhang, 2002]{zha:02}
Zhang, L.~C. (2002).
\newblock A method of weighting adjustment for survey data subject to
  nonignorable nonresponse.
\newblock {DACSEIS} research paper no. 2,
  http://w210.ub.uni-tuebingen.de/dbt/volltexte/2002/451.

\end{thebibliography}

\end{document}